\documentclass[aps,preprint,showpacs,preprintnumbers,amsmath,amssymb]{revtex4}
\usepackage{graphicx}
\begin{document}
\def\beq{\begin{equation}}
\def\eeq{\end{equation}}
\def\beqn{\begin{eqnarray}}
\def\eeqn{\end{eqnarray}}


\title{Interferences in the density of two Bose-Einstein condensates
consisting of identical or different atoms}

\author{L. S. Cederbaum$^{1}$, A. I. Streltsov$^{1}$, Y. B. Band$^{2}$, and O. E. Alon$^{1}$}
\address{$^1$Theoretical Chemistry, 
Heidelberg University, Im Neuenheimer Feld 229, 69120 Heidelberg \&
Max-Planck Institute for Nuclear Physics, Saupfercheckweg 1, 69117 Heidelberg, Germany}
\address{$^2$Department of Chemistry, Ben-Gurion University of the Negev,
Beer-Sheva 84105, Israel}


\begin{abstract}
The density of two {\it initially independent} condensates which
are allowed to expand and overlap can show interferences as a function of time
due to interparticle interaction.
Two situations are separately discussed and compared:
(1) all atoms are identical and
(2) each condensate consists of a different kind of atoms.
Illustrative examples are presented.
\end{abstract}
\maketitle

The investigation of interferences between particles is one of the
most basic tools to learn on the nature of quantum gases.
Interferences attracted much attention in particular in the case of
Bose-Einstein condensates (BECs) both from the theoretical and experimental sides,
see, e.g., \cite{r1,r2,r3,r4,r5,r6,r7}.
In a popular set up studied,
identical atoms are produced in two
traps which we may call the left and right traps and which are separated by a barrier.
By removing the traps and the barrier between them,
the atoms expand freely and can overlap.
In experiment, the photographs obtained show
spectacular interference fringes \cite{r1,r2}.

The interference of two parts of a single coherent condensate is by now well understood,
see, e.g., \cite{r4,r8,r9}.
On the other hand, relatively little is known on the interference of two
initially independent (i.e., fragmented) BECs, except for the case 
of non-interacting particles \cite{r3,r5,r6}.
Fragmented BECs can be produced using a barrier between the two traps 
which is so high and broad that tunneling between them is negligible.

In the available experiments, the atoms are prepared in a double-well trap
potential and it is not generally proven whether the atoms form a coherent
BEC, a fragmented BEC, or a combination thereof.  However, it is feasible
nowadays to produce in the lab two spatially separated, initially independent
BECs, see, e.g., \cite{r10}, and this allows for experiments with 
definitely fragmented BECs.
Apart from its importance as a fundamental problem, 
the solution of the problem of interference of two initially independent condensates 
is thus also of practical relevance.

In the scenario of two initially independent BECs the initial state 
of the many-body system before removing the traps reads
\beq\label{eq1}
 \left|\Psi\right> = \left(N_L! N_R!\right)^{-1/2} 
{(b_L^\dag)}^{N_L}{(b_R^\dag)}^{N_R} \left|vac\right>,
\ \ N_L+N_R=N, 
\eeq
where the $b_L^\dag$ and $b_R^\dag$ are the usual creation operators
for bosons in the left and right traps, respectively,
which contain definite numbers $N_L$ and $N_R$ of atoms in them.
After removing the traps, 
the state $\left|\Psi\right>$
is no longer an eigenstate of the system's Hamiltonian $H_0$
and expands in space as a function of time.
The time-dependent density, i.e., the expectation value of the density operator $\hat\rho(x)$
as a function of time becomes \cite{r7} 
\beq\label{eq2}
 \rho(x,t) \equiv \left<\Psi(t)\left|\hat\rho(x)\right|\Psi(t)\right> =
 N_L\left|\Phi_L(x,t)\right|^2 + N_R\left|\Phi_R(x,t)\right|^2, 
\eeq
where the $\Phi_{L,R}(x,t)$ are the single-atom states corresponding to
$b_{L,R}(t)=\exp(iH_0t)b_{L,R}\exp(-iH_0t)$.
Obviously, the density is a sum of the individual densities of the
two condensates and does not exhibit an interference term.

We would like to draw attention to the fact that the literature result 
(\ref{eq2}) has been obtained under the assumption that atoms belonging
to the two different BECs do not interact with each other.
Very recently it has been demonstrated that, in the presence of interaction,
the density $\rho(x,t)$ does show an interference term \cite{r12,r11}:
\beq\label{eq3}
 \rho(x,t)=\rho_{LL}(x,t)+\rho_{RR}(x,t)+\rho_{LR}(x,t),
\eeq
where $\rho_{LL}$ and $\rho_{RR}$ are the densities of the 
expanding separated BECs as if the two BECs do not communicate,
and $\rho_{LR}$ is the change of the density due to the interaction between them.
The terms contributing to $\rho_{LL}$ ($\rho_{RR}$) contain only
$b_L (b_R)$ and $b^\dag_L (b^\dag_R)$ operators, e.g.,
$b^\dag_L b^\dag_L b_L b_L$,
and those contributing to $\rho_{LR}$ contain only mixed products, e.g., $b^\dag_L b^\dag_R b_L b_R$.
The finding (\ref{eq3}) has many consequences.
In particular, the corresponding interference structures remain
after the statistical averaging over many experimental runs.
Of course, as $\rho(x,t)$ changes in time, 
the average must be carried out at the same value of $t$.

To derive (\ref{eq3}) the full Hamiltonian $H=H_0+V$ of the system after removing the
traps including the particle-particle interaction $V$ has been taken into account.
For the ease of presentation, we employ the widely used contact interaction
$V(x,x')=\lambda\delta(x-x')$,
where $\lambda$ is proportional to the s-wave scattering length \cite{r7,r13}.
Of course, any other interparticle interaction can be used as well.
As usual, $H_0$ describes the motion of the free atoms.
Starting from $H=H_0+V$ and the initial many-body
state (\ref{eq1}),
we have obtained the {\it exact} result for
$\rho(x,t)$ up to first order in the particle-particle interaction strength $\lambda$.
The corresponding expression is somewhat lengthy and is not given here,
but can be found in \cite{r11}.
Let us briefly mention properties of this result.
Clearly, the interference term $\rho_{LR}$ vanishes for $t\to 0$.
Furthermore, $\rho_{LR}(x,t)$ vanishes as expected if the atoms do not interact with
each other ($\lambda \to 0$).
The interference term $\rho_{LR}(x,t)$ is enhanced by the product $N_LN_R$
of the numbers of atoms in the two initial BECs.

The above discussions make clear that the interaction between the particles
gives rise to an interference term in the density of
two initially independent BECs of identical bosons.
Before presenting a numerical example we go one step further
and pose the question whether we can formulate a mean-field
theory which reproduces exactly the many-body small $\lambda$ result mentioned above.
Such a theory would open the door for real applications.
The usual mean-field theory leads to the well-known and widely used
Gross-Pitaevskii equation which reproduces exactly the density of BECs in a coherent 
state in the weak interaction limit \cite{r7,r13}.
Clearly, this equation is inapplicable to BECs in fragmented states (\ref{eq1}).
For fragmented states a more general multi-orbital
mean-field theory has been recently derived \cite{r14}.
In the present scenario two orbitals are involved and the respective
time-dependent mean-field [TDMF($2$)] takes on the appearance
(for the general derivation of TDMF, see \cite{r15}):
\beqn\label{eq4}
 i \dot\psi_L = {\mathcal P} \left[\hat h + \lambda(N_L-1)\left|\psi_L\right|^2 +
   2 \lambda N_R\left|\psi_R\right|^2\right] \psi_L, \nonumber \\
 i \dot\psi_R = {\mathcal P} \left[\hat h + \lambda(N_R-1)\left|\psi_R\right|^2 +
   2 \lambda N_L\left|\psi_L\right|^2\right] \psi_R \
\eeqn
where the initial conditions are 
$\psi_{L,R}(x,t=0)$.
$\hat h$ is the usual one-particle Hamiltonian
(in our scenario just the kinetic energy operator $-\frac{1}{2m}\frac{\partial^2}{\partial x^2}$) and
${\mathcal P}=1-\left|\psi_L\left>\right<\psi_L\right|-\left|\psi_R\left>\right<\psi_R\right|$
is a projector which ensures orthonormalization of the orbitals $\psi_L$ and $\psi_R$ \cite{r15}.
In TDMF($2$) the density can be expressed by 
$\rho(x,t)=N_L\left|\psi_L(x,t)\right|^2+N_R\left|\psi_R(x,t)\right|^2$.
Indeed, it can be shown \cite{r11} that the TDMF($2$) exactly reproduces the many-body result
in the weak interaction limit.

In the following we apply the TDMF($2$) theory (\ref{eq4}).
For coherent states the time-dependent Gross-Pitaevskii equation,
which is exact in the weak interaction limit,
has been demonstrated in many cases to be applicable for intermediate
and stronger interactions \cite{r7,r13}.
Similarly, there is reason to expect that for fragmented 
states the TDMF theory, which has been proven to be exact in the
weak interaction limit \cite{r11}, 
is applicable well beyond this limit.
We mention that TDMF($1$) is nothing but the 
time-dependent Gross-Pitaevskii equation.

\begin{figure}[ht] 
\includegraphics[width=11cm,angle=-0]{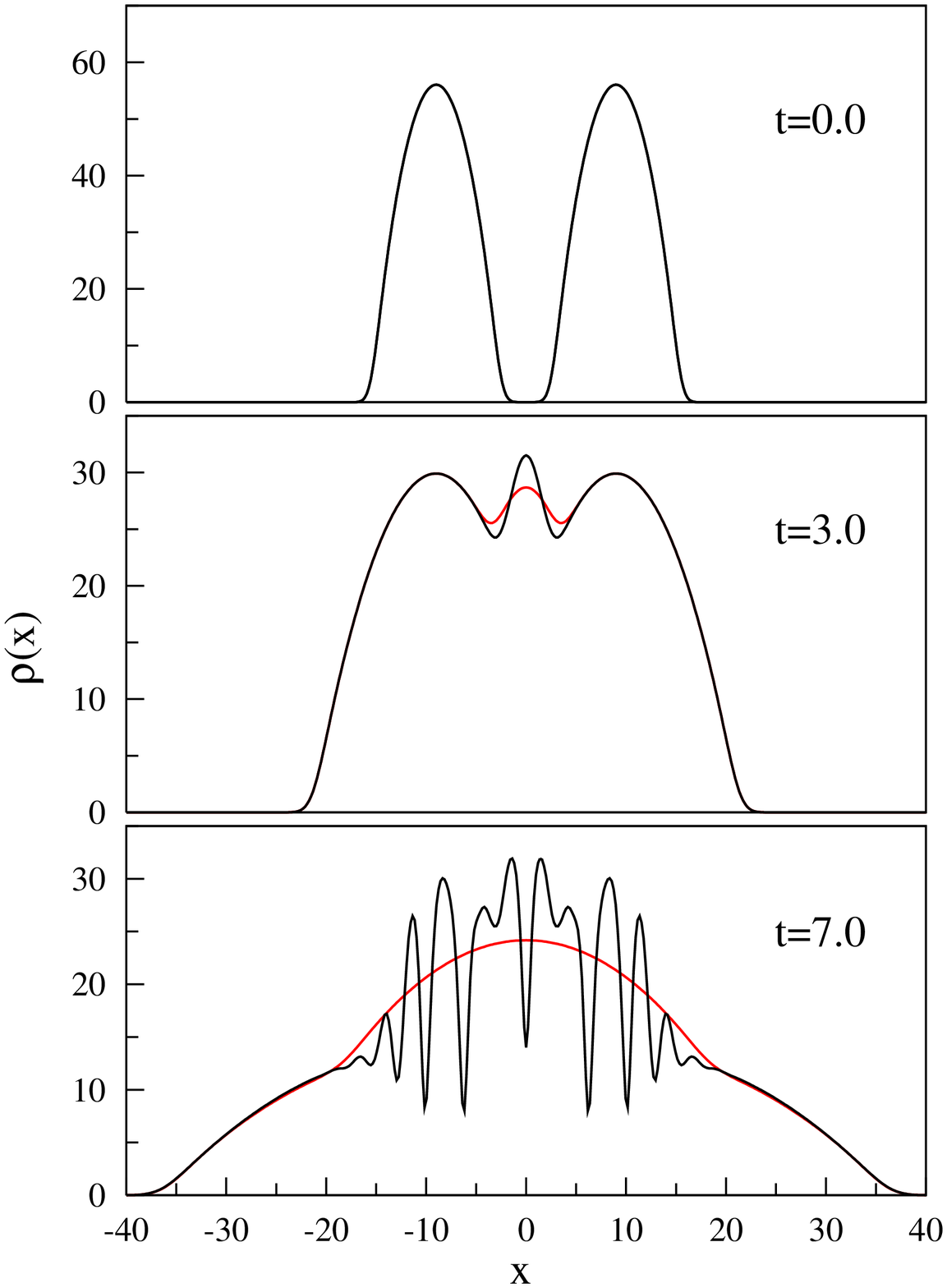}
\caption [kdv]{(Color online) The density $\rho(x,t)$ of two condensates
of $500$ atoms each for $\lambda=0.1$ as a function of time computed with TDMF($2$) (black)
compared to the density $\rho_{LL}+\rho_{RR}$ of
two BECs which do not interact with each other,
each computed with the Gross-Pitaevskii equation (red).
The quantities shown are dimensionless.
For more details see text.}
\label{F1}
\end{figure}

We consider harmonic traps centered at $\pm x_0$,
each containing a BEC with interaction $\lambda=0.1$.
At $t=0$ these traps are removed.
As initial conditions $\psi_{L,R}(x,0)$ 
we choose the respective
solutions of the stationary Gross-Pitaevskii equation at this $\lambda$
to account for the interaction when the harmonic traps are released.
In Fig.~\ref{F1} the density $\rho(x,t)$ computed using the TDMF($2$) equations 
is shown as a function of time.  
As seen in the figure, 
at $t=0$ the density consists of two separated distributions centered at $\pm x_0$.
The traps are removed at this time and the distributions start to broaden and to overlap.
At about $t=3$ one begins to see impact of the interference term in the density which
becomes strongly pronounced as time proceeds.

We see that the density of two initially independent condensates which are
allowed to overlap can show interference effects in the presence of interparticle interaction.
The physics of so called fragmented states, like the state in Eq.~(\ref{eq1}),
is generally very different from that of coherent states \cite{r16}.
Coherent states of condensates have been extensively studied,
mostly in the framework of the Gross-Pitaevskii equation \cite{r7,r13}.
A BEC in a coherent state can exhibit interference fringes
even in the absence of interaction \cite{r7,r8,r9,r17,r18}.
Take, for instance, the coherent state
$\left|\Psi^{coh}\right> = (N!)^{-1/2}{(b^\dag)}^{N}\left|vac\right>$
with $b^\dag=(b^\dag_L+b^\dag_R)/\sqrt{2}$.
This immediately leads to
$\rho^{coh}(x,t)=\frac{N}{2}\left|\Phi_L(x,t)+\Phi_R(x,t)\right|^2$ in the
absence of interaction between the atoms,
and hence to the interference term
$\rho_{LR}^{coh}(x,t)=N{\mathrm Re}\left(\Phi_L^\ast\Phi_R\right)$.
For expanding Gaussians with initial width $2a$ located at $\pm x_0$,
the oscillatory part of $\rho_{LR}^{coh}$ is simply given by
$\cos[K(t)x]$ with $K(t)=8x_0(t/m)/(a^4+4t^2/m^2)$.
This interference term is qualitatively different from that arising due to the 
interaction between the particles.
Another important difference between $\rho^{coh}_{LR}(x,t)$ and $\rho_{LR}(x,t)$
worth mentioning is that the former depends on the relative phase
between $\Phi_L$ and $\Phi_R$, 
while the latter does not depend on this phase.

Whether in an experiment the initial state is coherent or fragmented depends on the 
experimental conditions. 
It is beyond the scope of this work to argue whether or not the
initial state in the currently available experiments on interference is fragmented.
It is also not our intention to take side in the ongoing debate on whether these 
experiments detect the density or higher-order
correlation functions, 
although we tend to share the opinion of some researchers
see, e.g., \cite{r4,r8,r9,r17},
that the density is measured.
What we can state, is that if one
measures the density of two {\it freely} expanding initially independent BECs,
it will only show interferences in the presence of interaction.
This leads to the following proposal for an experiment
which makes use of the fact that nowadays one can vary the strength
of the interaction between the atoms \cite{r19,r20}.
Two measurements are necessary.
If the measurement with interaction shows interferences which
disappear upon measuring with the interaction turned off,
then (a) the initial state was a fragmented state 
and (b) the interaction is responsible for the interferences.

Until now the indistinguishability of the atoms has been
considered a precondition for interference effects.
In experiments on interferences one often starts with a single coherent BEC made of identical bosons
and produces two BECs by ramping up a barrier.
Interferences are then observed after removing the traps and the barrier.
Nearly all theoretical works on interferences in BECs rely on the
property of coherence of the whole system consisting of identical particles.
Even the two very recent works \cite{r12,r11} which discuss interferences
of identical independent BECs due to interaction
between the particles have assumed that indistinguishability is necessary.
We show below that this is unnecessary.

As above, we consider two initially independent BECs in an initial state like in (\ref{eq1}).
Now, however, each of these two BECs is made of a different kind of atoms.
For simplicity we call them ``left'' and ``right'' atoms,
and assign to them the creation operators $b_L^\dag$ and $b_R^\dag$ in (\ref{eq1}).
As usual, the Hamiltonian $H=H_0+V$ now contains three interactions terms
$V=V_L+V_R+V_{LR}$
accounting for the interaction between the ``left'' particles
[$V_L=\lambda_L\delta(x-x')$],
between the ``right'' particles
[$V_R=\lambda_R\delta(x-x')$],
and between the particles of both kinds
[$V_{LR}=\lambda_{LR}\delta(x-x')$], respectively.
Using the same basic techniques as in \cite{r11} for identical particles,
we obtain equation (\ref{eq3}) for the density also in the 
present case of {\it distinguishable} BECs.
The density contains a term $\rho_{LR}$ due to the interaction $V_{LR}$
between the particles of the two BECs.
Moreover, the analytic expression for $\rho$ in the weak interaction limit
is very similar to that in \cite{r11} for identical particles.

In contrast to the case of indistinguishable bosons,
where the time-dependent mean-field equations for fragmented condensates
have been derived very recently \cite{r15},
time-dependent mean-field equations for mixtures of different bosons are well known \cite{r21,r22,r23}.
For the present situation the latter read
\beqn\label{eq5}
 i \dot\psi_L = \left[\hat h_L + \lambda_L(N_L-1)\left|\psi_L\right|^2 +
     \lambda_{LR} N_R\left|\psi_R\right|^2\right] \psi_L, \nonumber \\
 i \dot\psi_R = \left[\hat h_R + \lambda_R(N_R-1)\left|\psi_R\right|^2 +
     \lambda_{LR} N_L\left|\psi_L\right|^2\right] \psi_R \
\eeqn
where for simplicity we have used the same nomenclature for
the orbitals $\psi_{L,R}(x,t)$ as for identical particles and the
kinetic energies $\hat h_{L,R} = -\frac{1}{2m_{L,R}}\frac{\partial^2}{\partial x^2}$
differ due to the possibly different masses of the left and right bosons.
It is not surprising that
(\ref{eq5}) reproduce the exact result for the density in the weak interaction limit.

It is illuminating to briefly compare equations (\ref{eq5}),
which we call time-dependent coupled Gross-Pitaevskii [TDCGP($2$)] equations,
to the TDMF($2$) for identical bosons (\ref{eq4}).
For identical bosons there are only a single mass $m$
and interaction strength $\lambda$,
and a factor $2$ appears in (\ref{eq4})
due to the exchange of identical bosons.
More importantly, 
TDMF($2$) maintains the orthogonality of the
orbitals $\psi_L$ and $\psi_R$
while TDCGP($2$) does not.

\begin{figure}[ht] 
\includegraphics[width=11cm,angle=-0]{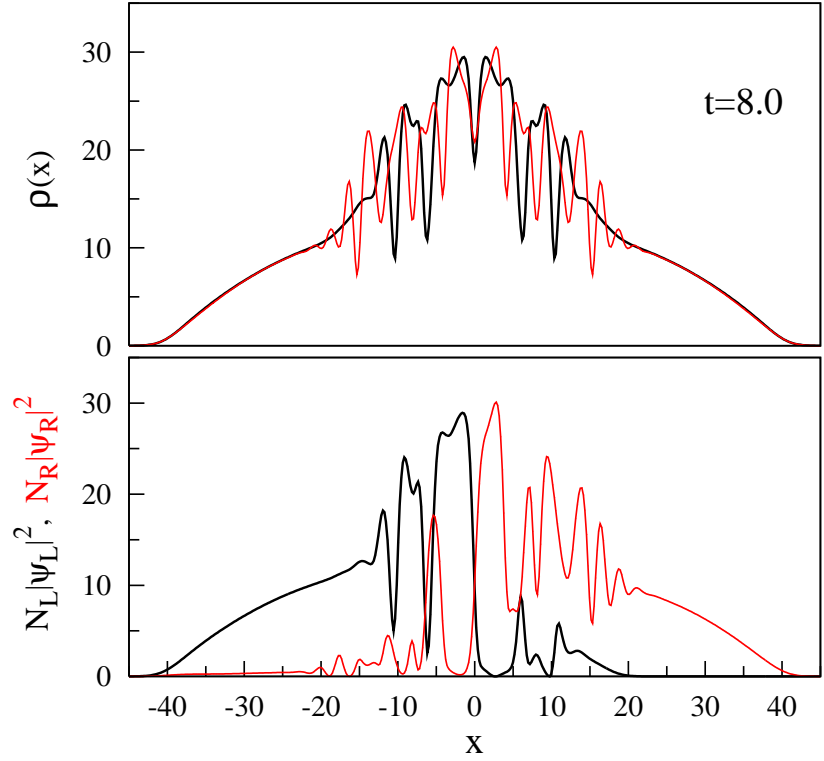}
\caption [kdv]{(Color online) The density of two condensates each made of $500$ atoms
of a different kind computed with
TDCGP($2$) for $\lambda_L=\lambda_R=0.1$ and $\lambda_{LR}=0.2$ at $t=8$ (black).
For comparison also the analogous result
for identical particles computed
with TDMF($2$) (for details see Fig.~\ref{F1}) is shown (red).
The lower panel shows the respective subdensities 
$N_L\left|\psi_L\right|^2$ for distinguishable atoms (black) and $N_R\left|\psi_R\right|^2$ 
for identical particles (red).}
\label{F2}
\end{figure}

To demonstrate that the density exhibits an oscillatory pattern also
for interacting distinguishable condensates,
we show a few examples computed via (\ref{eq5}).
For simplicity we put $m_L=m_R=m$.
In our first example we choose $\lambda_{LR}=2\lambda_L=2\lambda_R=0.2$ which leads
to the analogous scenario discussed in Fig.~\ref{F1}
for identical atoms.
At $t=0$ the initial density is as in Fig.~\ref{F1}.
At later times the density can evolve differently than in 
Fig.~\ref{F1} only because the orbitals $\psi_L$ and $\psi_R$ do not
have to be orthogonal to each other for distinguishable atoms.
In Fig.~\ref{F2} the density is shown for $t=8$ and compared with
the analogous density  obtained with (\ref{eq4}) for identical particles.
Both identical and distinguishable systems exhibit oscillatory structure
but their differences are substantial.
In the lower part of Fig.~\ref{F2} we show the individual subdensities
$N_L\left|\psi_L\right|^2$ and $N_R\left|\psi_R\right|^2$
at $t=8$ and remark that even a
moderate overlap $\left<\psi_L\left|\right.\psi_R\right>$ can have a 
considerable impact on the oscillatory pattern of the density.

\begin{figure}[ht] 
\includegraphics[width=11cm,angle=-0]{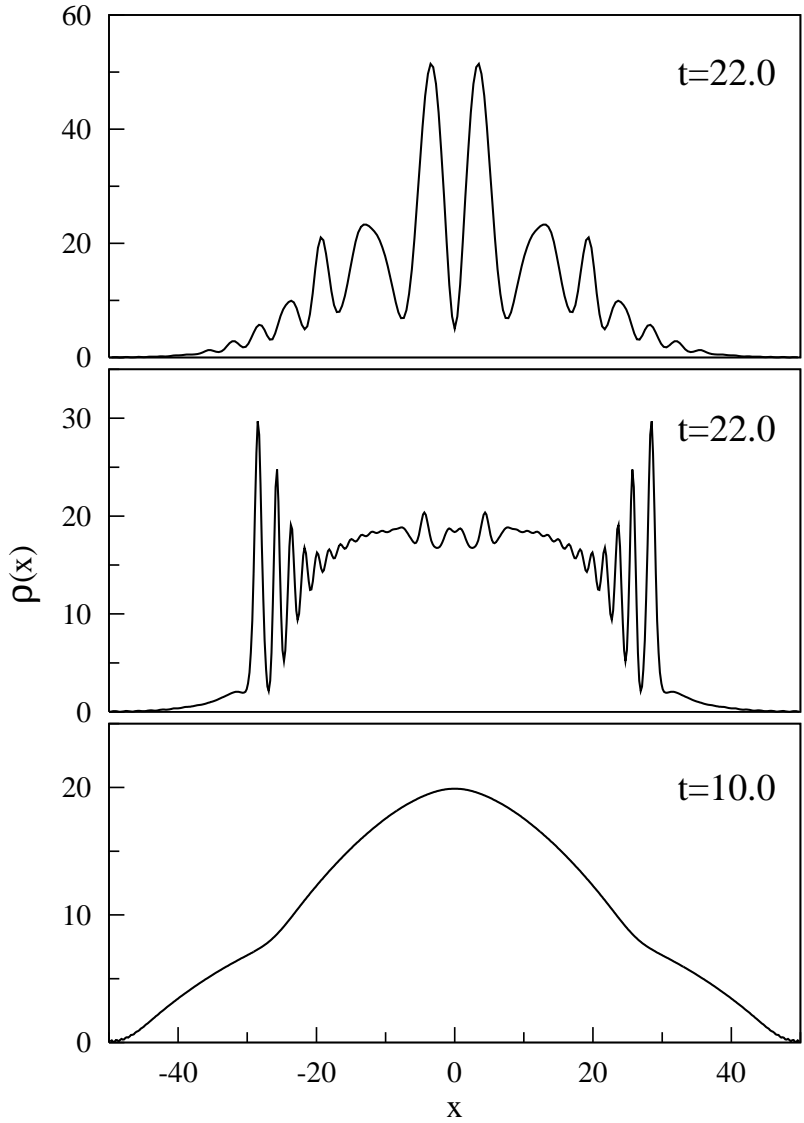}
\caption [kdv]{The density of two condensates
each made of $500$ atoms of a different kind.
Upper panel: repulsive interaction of the two condensates,
$\lambda_L=\lambda_R=0$, $\lambda_{LR}=+0.2$.
Middle panel: attractive interaction of the two condensates,
$\lambda_L=\lambda_R=0$, $\lambda_{LR}=-0.2$.
Lower panel: $\lambda_L=\lambda_R=\lambda_{LR}=0.1$.}
\label{F3}
\end{figure}

In our next example we choose $\lambda_L=\lambda_R=0$ and $\lambda_{LR}=0.2$, 
implying that the atoms in each of the left and right condensates do not interact,
but those belonging to different condensates do.
At $t=0$ we thus have two normalized Gaussians localized at the minima $\pm x_0$
of the harmonic traps.
After removing the traps these Gaussians expand and overlap,
and an oscillatory pattern develops.
The result is shown for $t=22$ in the upper panel of Fig.~\ref{F3}.
Until now the interactions studied were
repulsive and we address the question whether an
oscillatory structure can also arise
for attractive interactions.
To answer this question we investigate
the same scenario but with $\lambda_{LR}=-0.2$.
The result is depicted in the middle panel of Fig.~\ref{F3}.
Remarkably,
the oscillatory structure is even much more pronounced
than for the repulsive interaction.

In our last example we ask whether an oscillatory pattern 
always evolves when interactions are present.
We now choose $\lambda_L=\lambda_R=\lambda_{LR}=0.1$ and
find no oscillations in the density up to quite long times.
The density at $t=10$ is shown in the lower panel of Fig.~\ref{F3}.
This result can be understood from (\ref{eq5}).
Assuming $m_L=m_R$, $\lambda_L=\lambda_R=\lambda_{LR}$ and $N_L=N_R\gg 1$,
we notice that both equations (\ref{eq5}) are actually equivalent as they depend
only on the total density $\rho=N_L\left|\psi_L\right|^2+N_R\left|\psi_R\right|^2$.
The analogous situation is not possible for identical particles because of the factor
$2$ appearing in (\ref{eq4}) due to the exchange interaction.

The origin of the oscillations in the density is interference.
Commonly, one attributes a phase difference to the appearance of interferences.
This is particularly simple in the case of two parts
of a coherent condensate which show interference effects when overlapping.
One may attribute a phase to each of these parts.
As discussed above for fragmented condensates of identical particles,
the relative phase of the fragments is irrelevant in the context of interferences,
and this is, of course, also the case for condensates made of different
kinds of particles.
In these situations, interparticle interactions
are responsible for interference effects.
Here, we may speak of interaction-assisted self-interference.
Consider for simplicity two freely expanding, initially non-overlapping, different condensates with 
repulsive interaction between them.
In each of these condensates the orbital has a phase which
depends on $x$ and $t$.
As usual for a freely expanding isolated condensate,
its orbital and hence its phase are smoothly changing such that
no interference occurs.
Once the interacting condensates begin to overlap,
atoms are decelerated in the
overlapping region and the local phase changes there.
At later times the changed part of the orbital is superposed with other expanding parts 
of the same coherent condensate 
and this leads to interferences.
In the extreme case of infinitely strong repulsive interaction
between the two condensates, each of the condensates
is reflected from the other condensate as if it were an expanding hard wall. 
The interferences then arise from the superposition of the reflected and advancing parts
of the orbital.

The theory presented here is easily extendable to any kind of interparticle interaction. 
It is also easily extendable to the case where one does not let the
two BECs expand freely by removing the traps completely.
One may, e.g., remove only the barrier
and let the BECs expand in the new global trap.
Since the interference structures depend on the interaction,
a wealth of effects can be expected by varying the interaction,
the form of the individual traps
and the numbers $N_L$ and $N_R$ of the particles.
In the case of distinguishable particles also the
difference in masses $m_L$, $m_R$ and,
most importantly, the intra- and inter-condensate interactions
$\lambda_L$, $\lambda_R$ and $\lambda_{LR}$
enrich the possible range of interference phenomena.

\acknowledgments

\noindent
Financial support and collaboration within the 
German-Israeli DIP project of the German Federal Ministry of Education and Research
(BMBF) are gratefully acknowledged.

\end{document}